\documentclass{aa}

\usepackage{graphicx}
\usepackage{txfonts}
\begin{document}

   \title{The HARPS search for southern extra-solar planets.\thanks{Based on observations collected at ESO facilities
   under programs 082.C-0212, 085.C-0063, 086.C-0284, and 190.C-0027 (with the HARPS spectrograph at the ESO 3.6-m telescope, La Silla-Paranal Observatory). }}

  \subtitle{{XXXV. The interesting case of HD41248: stellar activity, no planets?}}

          \author{N.C. Santos\inst{1,2}
                 \and
                 A. Mortier\inst{1}
                 \and
                 J. P. Faria\inst{1,2}
                 \and
                 X. Dumusque\inst{3,4}
                 \and 
                 V. Zh. Adibekyan\inst{1}
                 \and 
                 E. Delgado-Mena\inst{1}
                 \and
                 P. Figueira\inst{1}
          	  \and
       	  L. Benamati\inst{1,2}
                 \and
                 I. Boisse\inst{8}
                 \and
                 D. Cunha\inst{1,2}
                 \and 
                 J. Gomes da Silva\inst{1,2}
                 \and
                 G. Lo Curto\inst{5}
                 \and
                 C. Lovis\inst{3}
                 \and
                 J. H. C. Martins\inst{1,2}
                 \and 
                 M. Mayor\inst{3}
       \and
         C. Melo\inst{5}
	  \and
	  M. Oshagh\inst{1,2}
          \and
          F. Pepe\inst{3}
          \and
          D. Queloz\inst{3,9}
       		\and
	    A. Santerne\inst{1}
	  \and
	  D. S\'egransan\inst{3}
	  \and
         A. Sozzetti\inst{7}
 	 \and
        S. G. Sousa\inst{1,2,6}
          \and
          S. Udry\inst{3}
          }
          

   \institute{     
         Centro de Astrof\'{\i}sica, Universidade do Porto, Rua das Estrelas, 4150-762 Porto, Portugal         
         \and
         Departamento de F\'{\i}sica e Astronomia, Faculdade de Ci\^encias, Universidade do Porto, Rua do Campo Alegre, 4169-007 Porto, Portugal
         \and
	Observatoire de Gen\`eve, Universit\'e de Gen\`eve, 51 ch. des Maillettes, CH-1290 Sauverny, Switzerland
	\and
	Harvard-Smithsonian Center for Astrophysics, 60 Garden Street, Cambridge, Massachusetts 02138, USA
	\and
	European Southern Observatory, Casilla 19001, Santiago, Chile
         \and
	Instituto de Astrof{\'\i}sica de Canarias, E-38200 La Laguna, Tenerife, Spain
	\and
	INAF - Osservatorio Astrofisico di Torino, Via Osservatorio 20, I-10025 Pino Torinese, Italy
	\and
	Aix Marseille Universit\'e, CNRS, LAM (Laboratoire d'Astrophysique de Marseille) UMR 7326, 13388, Marseille, France
	\and
	Institute of Astronomy, University of Cambridge, Madingley Road, Cambridge, CB3 0HA, UK
        }

   \date{Received XXX; accepted XXX}

 
  \abstract
   {The search for planets orbiting metal-poor stars is of uttermost importance for our understanding of the planet formation models.
   However, no dedicated searches have been conducted so far for very low mass planets orbiting such objects. Only a few
   cases of low mass planets orbiting metal-poor stars are thus known. Amongst these,
   HD\,41248 is a metal-poor, solar-type star on which a resonant pair of super-Earth like planets has been announced. 
   This detection was based on 62 radial velocity measurements obtained with the HARPS spectrograph (public data).}
   {In the present paper we present a new planet search program that is using the HARPS spectrograph to search
   for Neptunes and Super-Earths orbiting a sample of metal-poor FGK dwarfs.
   We then present a detailed analysis of an additional 162 radial velocity measurements of HD\,41248, obtained within this program, 
   with the goal of confirming the existence of the proposed planetary system.}
   {We analyzed the precise radial velocities, obtained with the HARPS spectrograph, together with several stellar activity diagnostics and line profile indicators. }
   {A careful analysis shows no evidence for the planetary system previously announced. One of the signals, with a period of $\sim$25\,days, is shown to be
   related to the rotational period of the star, and is clearly seen in some of the activity proxies. The remaining signal (P$\sim$18\,days)
   could not be convincingly retrieved in the new data set. }
  {We discuss possible causes for the complex (evolving) signals observed in the data of HD\,41248, proposing that they may
  be explained by the appearance and disappearance of active regions on the surface of a star with strong differential rotation,
  or by a combination of the sparse data sampling and active region evolution. }

\maketitle

\section{Introduction}
\label{sec:intro}

Precise spectroscopic studies of stars with {giant planets} show that their frequency is a strong function of the stellar metallicity. 
It is easier to find such a planet around a metal-rich star than around a metal-poor object \citep[][]{Gonzalez-1998,Santos-2001,Santos-2004b,Reid-2002,Fischer-2005,Sousa-2011}.
Several studies on solar neighborhood stars have shown 
that at least 25\% of stars with [Fe/H] above $+$0.3\,dex (twice the solar value) have an orbiting giant planet. This frequency
decreases to about 5\% for solar metallicity stars. This observational result is usually interpreted as due to a higher
probability of forming a giant planet core before the dissipation of the proto-planetary disk in a metal rich environment \citep[e.g.][]{Mordasini-2009a}.

A number of questions are still open, however, whose answer may have strong implications for planet formation models, especially in the metal-poor regime.
In the context of one of the HARPS surveys, a search for giant planets around a sample
of $\sim$100 metal-poor stars was conducted. Three new giant planet candidates were discovered,
and a fourth interesting candidate was announced \citep[][]{Santos-2007,Santos-2011}.
As expected, the results seem to confirm that metal-poor stars have a lower frequency of short-period giants \citep[see also][]{Sozzetti-2009},
and when these are found, they tend to have longer period orbits \citep[][]{Adibekyan-2013}.
Curiously, however, the results also suggest that the frequency of giant planets orbiting metal-poor stars may be higher than previously thought, at least for values of [Fe/H]$>-$0.7 \citep[][]{Mortier-2012}. 

Present numbers also indicate that the frequency of giant planets as a function of stellar metallicity may not be described
by a simple power-law \citep[as previously suggested for the metal-rich regime --][]{Johnson-2010}, and may be flat for metallicities below $-$0.1\,dex 
\citep[e.g.][]{Udry-2007,Mortier-2013a}. A tentative lower limit of the stellar
metallicity ($\sim$$-$0.7\,dex) below which no giant planets can be formed was also found \citep[e.g.][]{Mortier-2013a}. In brief, the giant planet formation 
efficiency in the metal-poor regime is still a matter of lively debate.
Since the metallicity is one of the most important ingredients controlling planet formation \citep[][]{Ida-2004b,Mordasini-2009a},
the answer to these issues is mandatory if we want to fully access the process of planet formation and evolution.

Additional information about the frequency of other types of planets (Neptune and super-Earth like) as a function of stellar metallicity is 
key in this discussion. 
In fact, contrarily to what one might expect, the known correlation between the presence of planets
and the stellar metallicity that exists for stars hosting {giant planets} 
does not seem to exist for stars hosting their lower mass planetary counterparts \citep[][]{Udry-2006,Sousa-2008}. Recent results have shown that stars with Neptune-mass planets 
have a rather flat metallicity distribution. Moreover, considering systems with 
only hot Neptunes (without any other Jupiter mass analog), though the number
is still small, the metallicity distribution becomes slightly metal-poor \citep[e.g.][]{Mayor-2011,Sousa-2011,Buchhave-2012}.

These observational facts are supported by theoretical work \citep[][]{Ida-2004b,Mordasini-2009a}, showing that 
{planets in the Neptune-mass regime should be common around stars with a wide range of metallicities}, while 
giant planets should be more common only around metal-rich stars. This can be interpreted as due to the fact that high metallicity proto-planetary
disks are able to form rocky/icy cores fast enough so that gas runaway accretion will lead to the formation of a giant planet before disk dissipation occurs. In turn, lower metallicity
disks will imply longer planet formation timescales, leading to a lower fraction of giant planets: cores don't grow fast enough to accrete gas
in large quantities before disk dissipation and thus remain ``Neptune" or ``Super-Earth" like.
However, given the still relatively small number of discovered low mass planets, and the reduced number of metal-poor stars 
surveyed {(no specific survey for low mass planets orbiting metal-poor stars has been carried out)}, 
it is still not possible to conclude on the frequency of low mass planets as a function of stellar metallicity.

In this paper we present a new project that makes use of precise HARPS radial velocities to search for Neptunes and Super-Earth planets orbiting a sample of metal-poor stars. 
The goals of the program and the sample are presented. We then turn our attention to the
case of \object{HD\,41248}, a metal-poor G dwarf from our sample that was recently announced to have a pair of resonant Super-Earths or Neptunes \citep[][]{Jenkins-2013}. 
Using the set with more than 200 precise radial velocities measurements together with different stellar activity diagnostics,
we explore the existence of the planets announced by \citet[][]{Jenkins-2013}. The results of this analysis are presented and discussed.

\section{The metal-poor survey}
\label{sec:survey}

To our knowledge, no specific radial velocity survey for Neptunes and Super-Earths orbiting a sample of 
low metallicity stars has been carried out. To tackle this issue, we started in October 2008 a dedicated program using the HARPS spectrograph at the 3.6-m ESO telescope (La Silla Paranal-Observatory, Chile). The first
set of observations, done in 3 different ESO periods between October 2008 and March 2011 (ESO programs 082.C-0212, 085.C-0063, and 086.C-0284) revealed several interesting candidates (see next section for the case presented here). However, despite the total granted 60 observing nights, the sparse time sampling of the observations did not allow us to conclude on the nature of any of the observed signals.

In order to address this problem, this initial observing program was granted an extra 80 nights over 3 years (starting in October 2012) within an ESO Large Program (190.C-0027). The goals of this program are twofold: first, to complete the search already started, and secondly to
confirm the very good candidates discovered in the previous runs. Once this program is finished, we expect to be able to derive the frequency of Neptunes and Super-Earths in the metal-poor regime 
and compare it with the published results for solar metallicity stars and with the model predictions \citep[e.g.][]{Mayor-2011}.
To achieve this goal, the idea is to obtain a number of points per star that is similar to the one obtained in the HARPS GTO survey for very low mass planets around solar neighborhood stars \citep[e.g.][]{Mayor-2011}, so that a similar detectability limit is reached.

The results of this survey will then allow us to compare the results and frequencies of Neptunes and Super-Earths with those 
obtained in the HARPS-Guarantee Time Observations (GTO) program to search for very low mass planets orbiting a sample of solar-neighborhood stars -- centered close to solar metallicity. Together, the surveys will set important constraints for the models of planet formation and evolution 
\citep[e.g.][]{Mordasini-2012}. Addressing this problem will help us to provide a proper estimate for the frequency of planets 
(including Earth-like planets) in our galaxy.

We would like to note that, as already widely known, the search for low mass planets, which induce very low amplitude signals in
the radial velocities, is a difficult and time-consuming process. This is very well illustrated by the huge
number of data points that was recently required to detect the Earth mass planet around $\alpha$\,Cen B \citep[][]{Dumusque-2012}. 
Further to this, a number of difficulties exist when dealing with the analysis of low amplitude signals.
For instance, stellar activity may induce false positive signals that can mimic the radial velocity signature of
a low mass planet \citep[e.g.][]{Forveille-2009}. Furthermore, recent results from radial velocity surveys have show that many of the low mass planets are in systems, where
several planets produce overlapping signals in the data, making the analysis even more complex \citep[e.g.][]{Lovis-2011}. The ubiquity of multi-planet systems
has also been demonstrated by the results of the Kepler mission \citep[e.g.][]{Batalha-2013}.

\subsection{Target selection and stellar properties}
\label{sec:sample}

The target list was chosen based on two sub-sets of the former HARPS GTO planet search program (completed in
2009). The first was a survey for {giant planets} orbiting metal-poor stars \citep[][]{Santos-2007}. 
The second was a program to search for {giant planets} orbiting a volume-limited sample
of FGK dwarfs \citep[][]{Naef-2007}. {Both sets of stars were surveyed
with a precision of $\sim$2-3\,m\,s$^{-1}$, clearly insufficient to 
allow for the detection of Neptune-like planets (the observing strategy and frequency of measurements 
was also not adequate for this goal)}. 

Merging these two HARPS samples, we took all stars that met the following criteria:

\begin{itemize}
\item Not known to harbor low mass planets;
\item Metallicities below $-$0.4\,dex \citep[derived from HARPS CCF -- see e.g.][]{Santos-2002a};
\item Chromospherically quiet ($\log{R'_{HK}}<-4.8$, as measured from HARPS spectra);
\item Present radial-velocity variations with a dispersion below 10\,m\,s$^{-1}$ (higher dispersions may imply the presence of higher mass planets);
\item Brighter than V=9.5 (to allow a photon noise precision of 1\,m\,s$^{-1}$ after 900s).
\end{itemize}

The previous information gathered in both surveys was thus sufficient to allow for the definition of a
good sample of 109 metal-deficient stars ($-$1.5$<$[Fe/H]$<-$0.4) that are suitable targets for the detection of very low mass planets (Neptunes or Super-Earths) -- Table\,\ref{tab:sample}. To these we added the three long period planet host stars presented in \citet[][]{Santos-2011}, whose planets
were discovered in the context of the HARPS GTO program to search for giant planets orbiting metal-poor
stars (\object{HD171028}, \object{HD181720}, and \object{HD190984}), as well as a fourth long period planet host candidate (\object{HD107094}). The goal 
is to search for very low mass planets orbiting these stars.

\begin{table}[t!]
\caption{\label{tab:sample}. List of stars in the sample, their V magnitudes, and coordinates (2000.0 equinox).}
\centering
\begin{tabular}{lccc}
\hline\hline
Star & V &RA(J2000)&Dec(J2000)\\
\hline
HD224817 & 8.41 & 00:00:58.2& $-$11:49:25\\
HD208 & 8.23 & 00:06:53.9& $-$03:37:34\\
HD967 & 8.36 & 00:14:04.5& $-$11:18:42\\
(...) & (...) & (...) & (...)\\\hline
\end{tabular}
\end{table}

\begin{figure}[t!]
\resizebox{8cm}{!}{\includegraphics[bb=20 170 580 700]{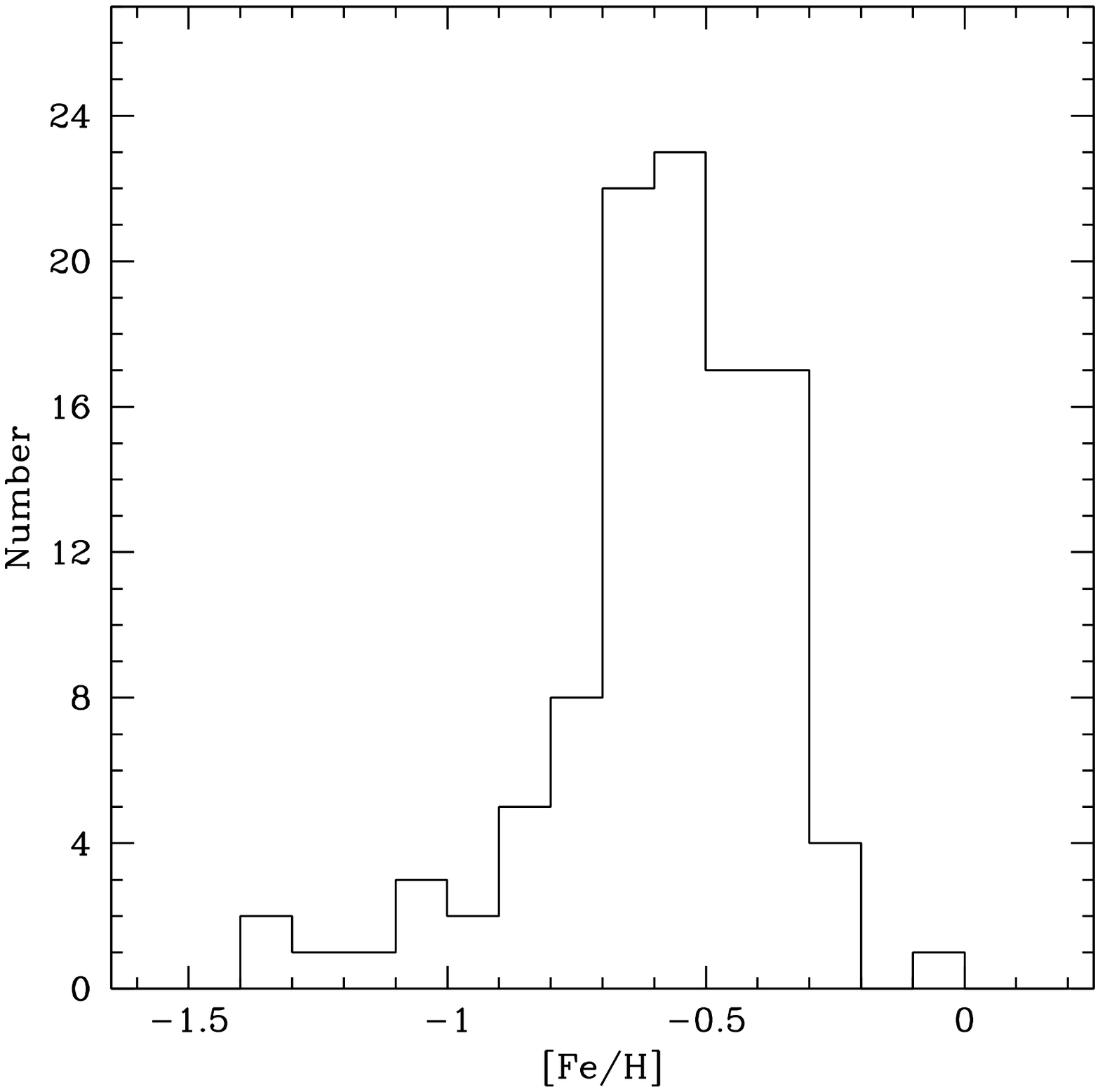}}
\resizebox{8cm}{!}{\includegraphics[bb=20 170 580 700]{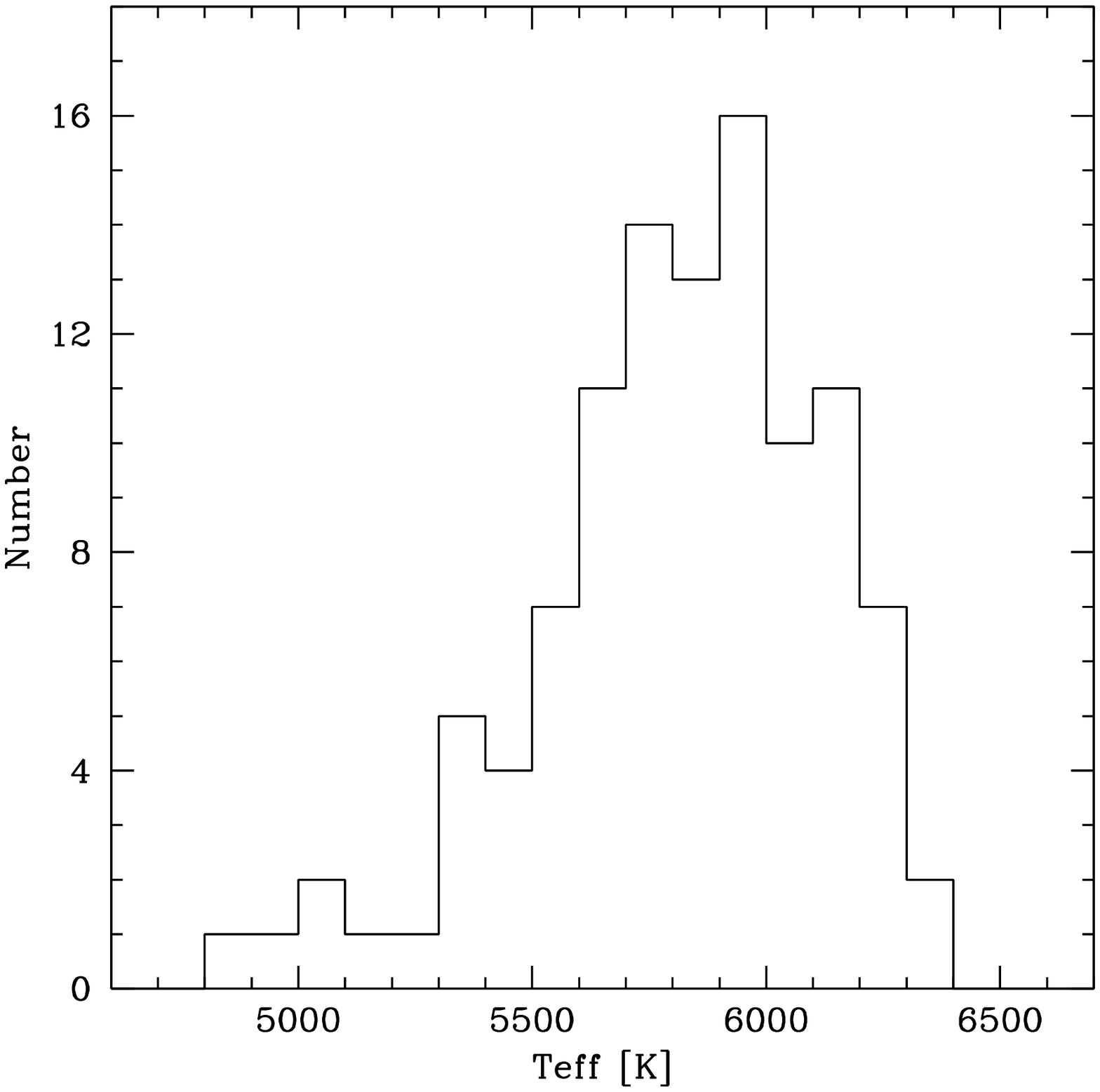}}
\caption{Metallicity (top) and effective temperature (bottom) distributions for the stars in our sample.}
\label{fig:param}
\end{figure}

Stellar parameters for 106 out of the 109 targets were derived from a set of high resolution HARPS spectra taken during the 
HARPS-GTO program\footnote{Exceptions are \object{HD\,196877},  \object{HD\,211532},  and  \object{HD\,304636}.}. The values were presented in \citet[][]{Sousa-2011,Sousa-2011b}. In Fig.\,\ref{fig:param} we present the distributions of [Fe/H] and effective temperature. All the stars in the sample have metallicities lower than solar, with an average [Fe/H] of $-$0.58. However, since these parameters were derived after the sample has been defined\footnote{As mentioned above, the definition was done using values derived from a calibration of the HARPS-CCF, and not from a detailed spectroscopic analysis}, not all stars have a metallicity below $-$0.4\,dex: 21 stars have spectroscopic [Fe/H] values higher than this value, though lower than $-0.22$\,dex. One single ``outlier'' was found to have almost solar metallicity (\object{HD\,144589}, [Fe/H]=$-$0.05).

\section{HD41248}

In a recent paper, \citet[][]{Jenkins-2013} used the first 62 (public) radial velocity measurements of the star HD\,41248 to announce the detection of
a system of two super-Earth or Neptune mass planets, with orbital periods of $\sim$18.36 and 25.65 days. While the second of these signals 
was not very conspicuous, the first is clear on the dataset they analyzed.

The 18-day period signal had been spotted by us in 2010. However, since its value was close to the expected rotational period 
of the star as computed from the activity level \citep[see Sect.\,\ref{sec:star}, as well][]{Jenkins-2013}, and because a possible peak at the same period
was also seen in the Bisector Inverse Slope (BIS) of the HARPS Cross Correlation Function (CCF) -- Fig.\,\ref{fig:perioset1} -- we decided that it would be wise to gather a new batch of observations before announcing the putative planet. In the following we will present the results of the analysis of the whole date set gathered in the programs presented above.

\subsection{Stellar properties}
\label{sec:star}

\object{HD\,41248} (HIP\,28460) is a V=8.82 magnitude G2 dwarf in the southern constellation Pictor. According to the new Hipparcos catalog reduction \citep[][]{vanLeeuwen-2007}, it has a parallax of 19.22$\pm$0.79\,mas, which puts it at a distance of 52$\pm$2 parsec from the Sun. 
\citet[][]{Sousa-2011} derived precise stellar parameters for this star using a set of high resolution and high S/N spectra obtained during the HARPS-GTO program. The resulting values are T$_{\mathrm{eff}}$=5713$\pm$21\,K, $\log{g}$=4.49$\pm$0.05\,dex, and [Fe/H]=$-$0.37$\pm$0.01\,dex\footnote{These errors are merely the internal uncertainties. For the surface gravity and metallicity we adopt more realistic uncertainties of 0.05\,dex (reasonable given the proximity of the effective temperature to the solar one).}. These values are very similar to the ones listed by \citet[][]{Jenkins-2013}: T$_{\mathrm{eff}}$=5713$\pm$50\,K, $\log{g}$=4.48$\pm$0.10\,dex, and [Fe/H]=$-$0.43$\pm$0.10\,dex. Compatible values for the effective temperature are also listed in the PASTEL catalogue \citep[][]{Soubiran-2010}: 
\citet[][]{Masana-2006} derived T$_{\mathrm{eff}}$=5827\,K, while \citet[][]{Casagrande-2011} obtained T$_{\mathrm{eff}}$=5927\,K.
 
An estimate for the mass and radius of \object{HD\,41248} can be obtained using the calibration in \citet[][]{Torres-2010}. The value and its uncertainty was derived using a Montecarlo approach, where random values of effective temperature, surface gravity, and metallicity as derived by \citet[][]{Sousa-2011} were drawn taking into account the (gaussian) uncertainties. Final values of 0.94$\pm$0.02\,M$_\odot$ and 0.92$\pm$0.06\,R$_\odot$ were derived for the mass and radius, respectively.
Using this value for the stellar mass, the effective temperature derived by Sousa et al., the parallax, the visual magnitude, and the bolometric correction of $-$0.09 as derived from the calibration of \citet[][]{Flower-1996}, we derive an astrometric surface gravity of 4.56\,dex \citep[see Eqn.\,1 in][]{Santos-2004b}, very similar to the spectroscopic value. These are typical stellar parameters for a G2, moderately metal-poor dwarf.

The analysis of the HARPS spectra (see below) also allows us to derive the stellar activity level of the star, using the \ion{Ca}{ii} H and K lines \citep[][]{Lovis-2011b}. The average value over the $\sim$10 years of measurements is <$\log{R'_{HK}}$>=$-$4.90, with the values ranging from $-$5.20 to $-$4.79.
These values are typical for a solar-like activity star in the low activity part of the Vaughan-Preston gap \citep[][]{Vaughan-1980}.
On its side, the observed value can be used to derive an estimate for the rotational period of the star. Using the calibrations of \citet[][]{Noyes-1984} and \citet[][]{Mamajek-2008} we obtain values for the rotational period of 19.8$\pm$3.6 and 20.1$\pm$3.0\,days\footnote{Uncertainties are computed from the rms of the $\log{R'_{HK}}$ values.}, respectively. Finally, from the FWHM of the HARPS cross correlation function we could estimate a value of 1.0\,km\,s$^{-1}$ for the projected rotational velocity of the star \citep[see e.g.][]{Santos-2002a}, a value slightly lower than the 2.4\,km\,s$^{-1}$ listed by \citet[][]{Jenkins-2013}.

\begin{table}[t!]
\caption{Stellar parameters for HD41248.}
\label{tab:parameters}
\begin{tabular}{lc}
\hline\hline
\noalign{\smallskip}
Parameter  			& \\
\hline
Spectral~type			& G2V\\
$m_v$				& 8.82\\
$B-V$				& 0.62 \\
Parallax [mas]			& 19.11$\pm$0.71\\
Distance~[pc]			& 52$\pm$2\\
M$_v$				& 5.23\\
L [L$_{\odot}$]			& 0.70\\
$\log{R'_{\rm HK}}$		&$-$4.90\\
P$_{Rot}$ [days]		&20$\pm$3\\
$v\,\sin{i}$~[km~s$^{-1}$]	&1.0\\
$T_{\rm eff}$~[K]  		& 5713$\pm$21 \\
$\log{g}$				& 4.49$\pm$0.05\\
${\rm [Fe/H]}$			&$-$0.37$\pm$0.05 \\
Mass~$[M_{\odot}]$		& 0.94$\pm$0.02 \\
Radius~$[R_{\odot}]$	& 0.92$\pm$0.06 \\
\hline
\noalign{\smallskip}
\end{tabular}
\end{table}

\object{HD\,41248} is a thin-disk star both in terms of kinematics and chemistry. With a value of [$\alpha$/Fe] = 0.05 dex \citep[][]{Adibekyan-2012b}, the star does not show any $\alpha$-element enhancement, a characteristic that is used to distinguish thin and thick disk stars at that metallicity \citep[][]{Fuhrmann-1998,Bensby-2003,Adibekyan-2012b}. Its oxygen-to-iron abundance ratio, derived using the OI\,6300\AA\ line is [O/Fe] = 0.11\,dex, also in agreement with the results for other $\alpha$-elements. {Note also that the $\alpha$-enhancement has been shown to correlate with the presence of planets in the metal-poor regime \citep[][]{Haywood-2008,Adibekyan-2012}.}

The galactic space velocity components of the star ($U{}_{\mathrm{LSR}}$ = -2, $V{}_{\mathrm{LSR}}$ = -6, and $W{}_{\mathrm{LSR}}$ = 34 km s$^{-1}$) also suggest a thin-disk origin with a probability of $\sim$95\% \citep[][]{Adibekyan-2012b}. \object{HD\,41248} has a low Galactic orbital eccentricity (0.04) and low Z$_{max}$\footnote{The maximum vertical distance the stars can reach above/below the Galactic plane.} of about 0.6 kpc \citep[][]{Casagrande-2011}.
Finally, it shows a Li abundance of 1.56$\pm$0.10 \citep[][]{Delgado-Mena-2014}. This value is typical for a star of its effective temperature, and does not reflect any particularly strong Li depletion as often found in planet-host stars of similar temperature \citep[see e.g.][]{Delgado-Mena-2014}.

\subsection{Radial velocities}
\label{sec:rv}

Between October 2003 and December 2013, a total of 223 radial velocity measurements were obtained of HD\,41248
using the HARPS spectrograph at the 3.6-m ESO telescope (La Silla-Paranal Observatory). The simultaneous calibration
mode was used. Starting in March 2013, the simultaneous calibration was done using the available Fabry-Perot system,
while before this date the ThAr lamp was used in this process. The average signal-to-noise of the HARPS
spectra in order 60 ($\sim$6200\AA) is 93, with values ranging from around 20 up to 150.

An analysis of the HARPS spectra allows us to exclude problematic measurements a priori (before the
radial velocity analysis). Such situation include e.g. measurements with very low S/N or spectra with an abnormal
blue-to-red flux ratio. These cases are usually related with nights when the transmissions was particularly bad (e.g. the presence of cirrus)
or when observations are done at high airmass values. Two of the measurements in our data set (JD=55304.518017 and 56409.495511) 
were excluded based on these criteria. In all the analysis presented in this paper we made use of the remaining 221 data points.

The radial velocities (RVs) were derived using the HARPS pipeline (version 3.7) making use of the weighted Cross-Correlation technique,
and using a cross-correlation mask optimized for a G2 dwarf (the same spectral type as HD\,41248). The average error of the 
RVs is 1.4 m\,s$^{-1}$. This value includes the photon noise, the calibration noise, and the uncertainty in the measurement of the instrumental drift.
In all subsequent analysis, an error of 70 cm\,s$^{-1}$ was further quadratically added to this uncertainty, to take into account
other possible sources of noise including instrumental, atmospheric, and stellar contaminants, as well as stellar jitter \citep[see e.g.][]{Pepe-2011}. 
The addition of this white-noise will not introduce any inexistent signals in the data.

As presented in Sect.\,\ref{sec:sample}, the first set of RV data points obtained for this star was gathered in the context of a sub-survey of the
HARPS GTO program. The goal of this sub-sample was to search for giant planets, and a corresponding
strategy, in terms of precision, was adopted. As such,
the error bars in a large fraction of the first dataset are significantly higher than the ones found in later measurements. Since October 2008 
(when the large program started) the measurements were obtained with a completely different strategy. 
Exposure times were set at a minimum of 15 minutes in order to average-out the noise coming from stellar
oscillations \citep[e.g.][]{Santos-2004a}.
Starting in October 2012 we also decided to obtain, whenever possible, more than 1 spectrum of the star in a given night, separated by 
several hours. This strategy was used to minimize sources of stellar noise such as stellar oscillations and granulation and has proven to be very efficient when 
searching for extremely low amplitude RV signals \citep[][]{Pepe-2010,Dumusque-2010}. Since the periodic signals we will be analyzing in this paper are much longer than one night, we used the 
nightly binned data in our analysis. This implies that ``only'' a total of 156 separate data points (in 156 different nights) are considered\footnote{These RV measurements are
published as an electronic table}.

\begin{figure}[t!]
\resizebox{\hsize}{!}{\includegraphics[bb=10 150 580 720]{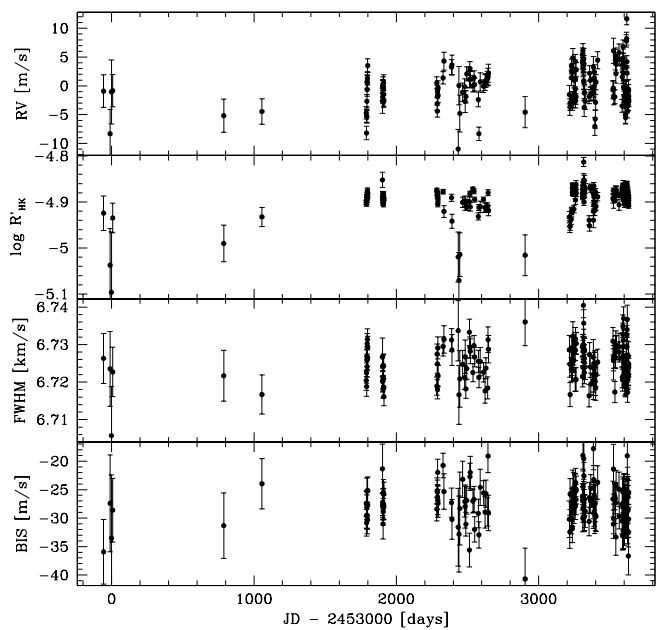}}
\caption{Time series of the radial velocity, stellar activity, FWHM, and BIS.}
\label{fig:alldata}
\end{figure}

The HARPS pipeline also derives values for other parameters such as the Bisector Inverse Slope \citep[BIS --][]{Queloz-2001}, the Cross-Correlation Function (CCF) parameters FWHM and Contrast, as well as the activity level of the star using the \ion{Ca}{ii} K and K lines ($\log{R'_{HK}}$). To these, we have also separately computed (using the software described in Appendix\,\ref{appendix}) a number of alternative line profile variation indicators as defined in \citet[][V$_{span}$]{Boisse-2011}, \citet[][biGauss]{Nardetto-2006}, and \citet[][V$_{asy}$, BIS$+$, and BIS$-$]{Figueira-2013}. These indices are used for the analysis and interpretation of the observed radial velocity signals. 

In Fig\,\ref{fig:alldata} we plot the radial velocity time series, together with the derived values for the activity level and the CCF parameters FWHM and BIS. 
A simple look at the plots shows that the RV values show a slightly increasing trend with time. No clear trend is seen for the $\log{R'_{HK}}$ activity index,
the FWHM, or the BIS, which suggests that this drift is not related to the variation of activity level along the magnetic cycle of the star \citep[][]{Santos-2010a,Lovis-2011b,Dumusque-2011}.


\subsection{Keplerian fitting}
\label{sec:keplerians}

To test if the signals detected by \citet[][]{Jenkins-2013} are still present in the data after including the additional RVs, we decided as a first approach  
to use the yorbit algorithm (S\'egransan et al., in prep.) to fit the whole data set with a model composed of
2 Keplerian functions and one linear trend. Yorbit uses an hybrid method based on a fast linear algorithm (Levenberg-Marquardt) 
and genetic operators (breeding, mutations, cross-over), and has been optimized to explore the parameter space when doing Keplerian fitting
of radial velocity data sets. Since the first goal was to explore the existence of the signals announced by
Jenkins et al., we chose to explore only the solutions with periods between 16 and 20 days (for the first planet), and between
24 and 28 days (for the second planet). 

\begin{figure}[t!]
\resizebox{\hsize}{!}{\includegraphics[]{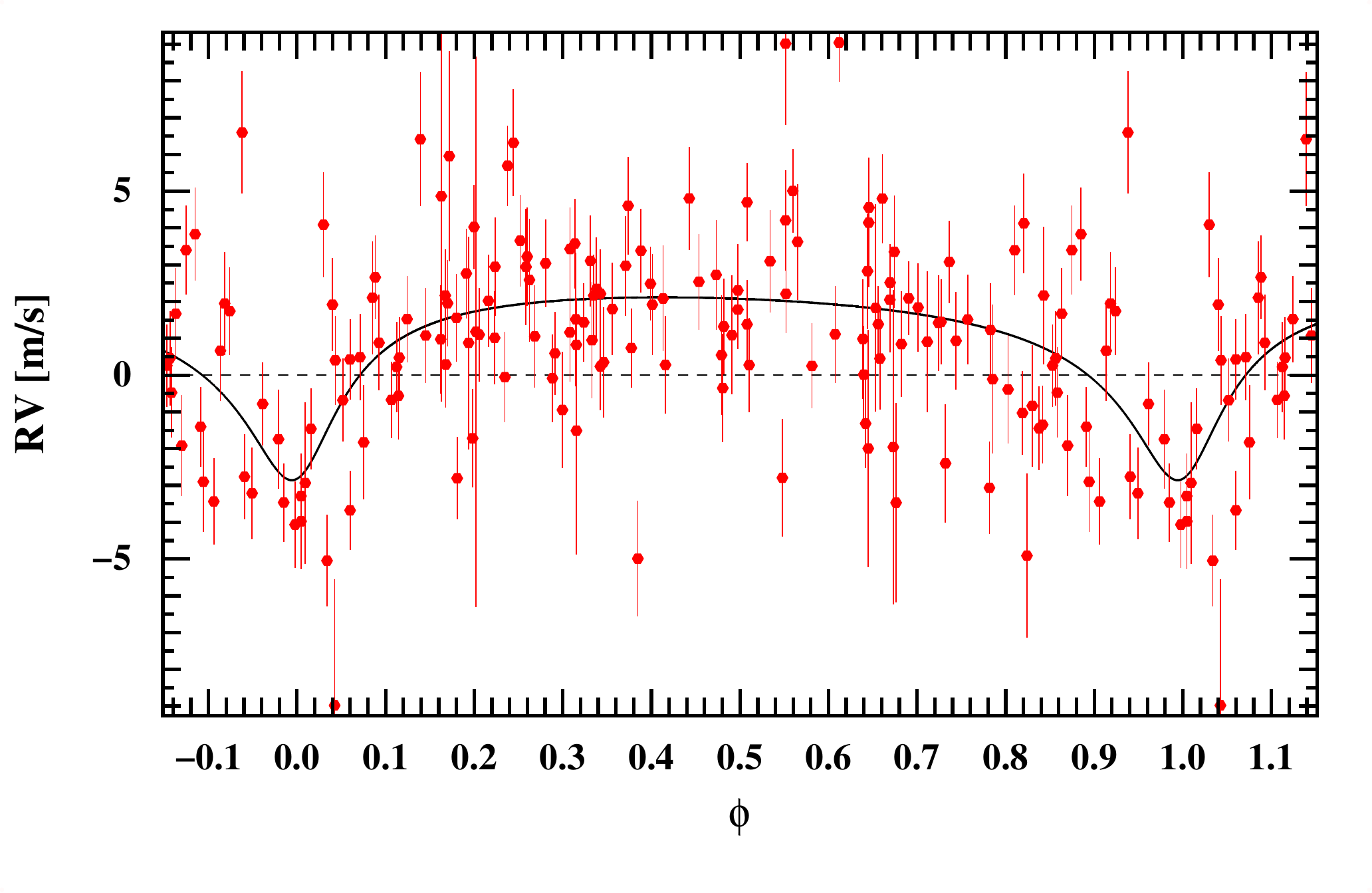}}
\resizebox{\hsize}{!}{\includegraphics[]{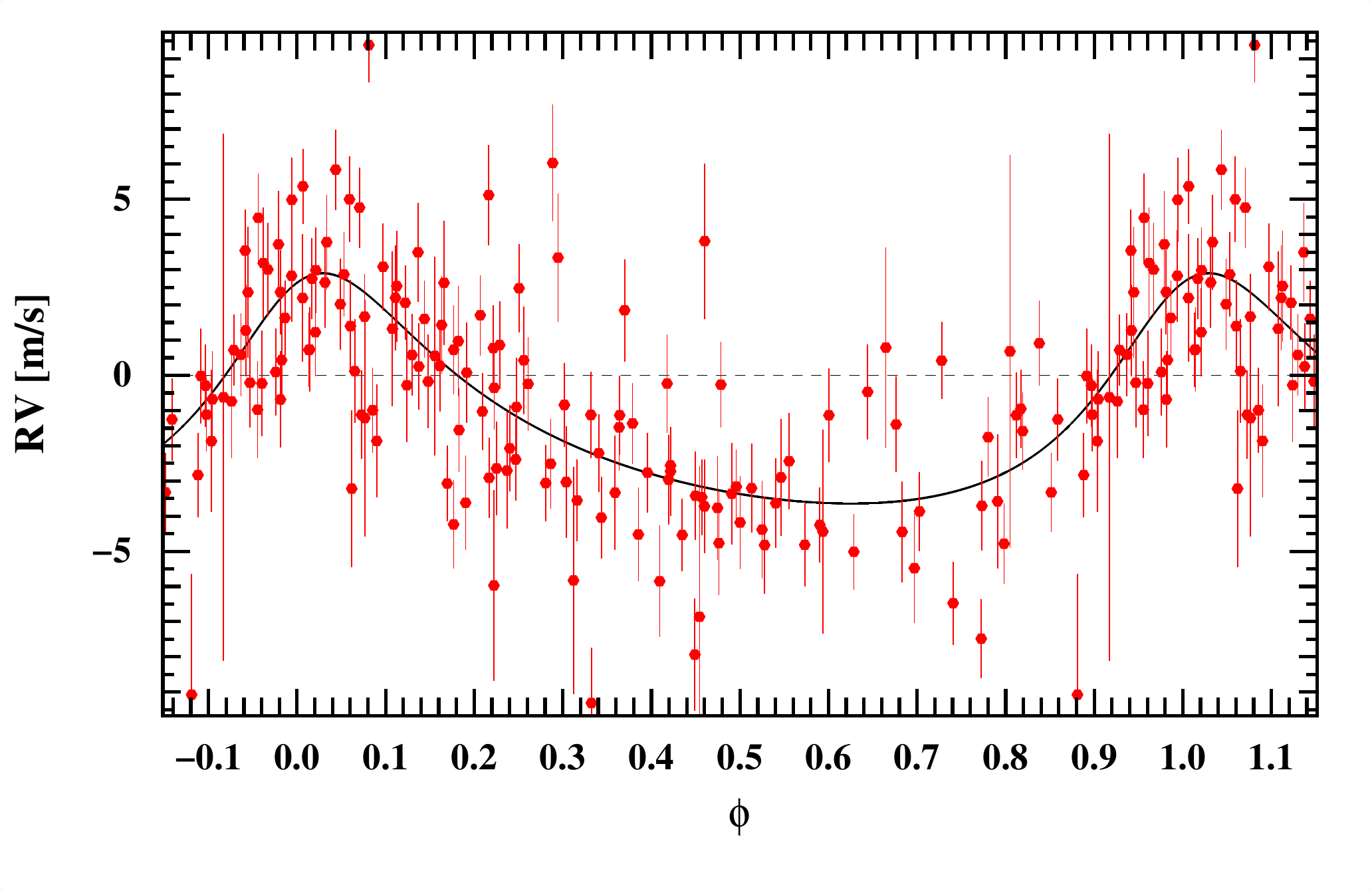}}
\caption{Phase folded radial velocities of HD\,41248 with the best fit Keplerian functions at $\sim$18-days (top) and $\sim$25\,days (bottom).}
\label{fig:keplerians}
\end{figure}

The phase folded, best fit Keplerian solutions are presented in Fig.\,\ref{fig:keplerians}. The final solutions
converged to orbital periods of 18.336$\pm$0.006 and 25.623$\pm$0.010\,days, with eccentricities of 0.54$\pm$0.09 and 0.36$\pm$0.07 and semi-amplitudes 
of 2.46$\pm$0.41 and 3.32$\pm$0.26 m\,s$^{-1}$, respectively. If caused by the presence of planets, and assuming a stellar mass of 0.94\,M$_{\odot}$, these solutions correspond to the
signal induced by Super-Earth or Neptune like planets with masses of 8.2 and 13.7\,M$_\oplus$, respectively. 
The residuals to the fit show an rms of 2.15 m\,s$^{-1}$, clearly above the average error bar of the measurements (1.4 m\,s$^{-1}$).
Some structure is also present in the residuals: a Generalized Lomb Scargle periodogram \citep[GLS,][]{Zechmeister-2009b} shows some power at $\sim$30 days, though not statistically significant (the computed False Alarm Probability is around 5\%). Given the complexity of the observed signals (see discussion below),
in the present paper we will not discuss the nature of this signal in detail\footnote{As we will see below, a peak at this period is clearly also seen in the FWHM, BIS$-$, and V$_{asy}$ periodogram of set\,\#1 (Fig.\,\ref{fig:perioset1}).}.

The 25-day period fit presented in Fig.\,\ref{fig:keplerians} looks perfectly reasonable. However, the 18\,day signal is
visually not ``convincing'', as it owes its shape mostly to a few points near phase 0 (or 1).
But even if it was credible, the fact that visually these solutions could be acceptable does not, of course, confirm the existence of planets orbiting
HD\,41248. For instance, several cases have shown that stable active regions can be present in the photospheres of solar-type stars \citep[e.g.][]{Queloz-2001,Forveille-2009,Figueira-2010b}. These may
produce RV signals that mimic the ones expected from real planetary systems.

A simple comparison of the fitted signals with the ones presented in \citet[][]{Jenkins-2013} shows that the periods found are consistent.
The eccentricities, however, are significantly higher than the ones  (close to zero) presented by these authors. The amplitude of the 25\,day period signal
is also significantly above the maximum value listed by Jenkins et al. (2.97 m\,s$^{-1}$). Imposing circular orbits decreases the amplitudes to 1.99 and 2.99 m\,s$^{-1}$, respectively,
but produce a slightly worse fit with an rms of 2.26 m\,s$^{-1}$. In any case, these values suggest that at least the 25-day period signal has
evolved in amplitude over time.

\subsubsection{Bayesian analysis with Keplerian functions}
\label{sec:bayes}

In complement, we also performed a Bayesian analysis of the whole data set following the methodology done e.g. by \citet[][]{Gregory-2011}. In this process
we used large and uninformative priors, except for the orbital eccentricity for which we choose a Beta distribution as suggested by \citet{Kipping-2013}. 
We also assumed here that the data can be modeled by a series of Keplerian orbits and a linear drift.
We ran a large number of chains using the Markov Chain Monte Carlo (MCMC) algorithm implemented into the \texttt{PASTIS} software 
\citep[][]{Diaz-2014}. We point to this paper for more details on all the process. We then compute the Bayes factor of models with $n+1$ Keplerian orbits against models with $n$ Keplerian orbits by estimating the evidence of each model using the Truncated Posterior Mixture as described by \citet{Tuomi-2012}.


We found that the data can be modeled by up to five significant Keplerians with periods respectively of $P_{1} = 25.628  \pm 0.011$\, days, $P_{2} = 18.349  \pm  0.012$\, days, $P_{3} = 30.715  \pm 0.031$\, days, $P_{4} = 12.6291  \pm  0.0034$\, days, and $P_{5} = 8.8^{_{+1.2}}_{^{-1.7}}$\, days. All these Keplerian orbits are found to have significant eccentricities, except for the ones at $\sim$18 and $\sim$8.8\,days. The two first Keplerian orbits are compatible with the ones reported by \citet[][]{Jenkins-2013}, while $P_{4}$ and $P_{5}$ are found to be really close to $P_{1}/2$ and $P_{1}/3$. 

As we will see below, we will conclude that the $P_{4}$ and $P_{5}$ are the harmonic of the stellar activity signal which has a main period $P_{1} \approx25.6$\, days. We also find
strong indications that the third Keplerian orbit, with $P_{3} \approx 30.7$\,days, is related to stellar activity.

\subsection{Analyzing the periodograms}
\label{sec:periodograms}

\begin{figure}[t!]
\resizebox{\hsize}{!}{\includegraphics[]{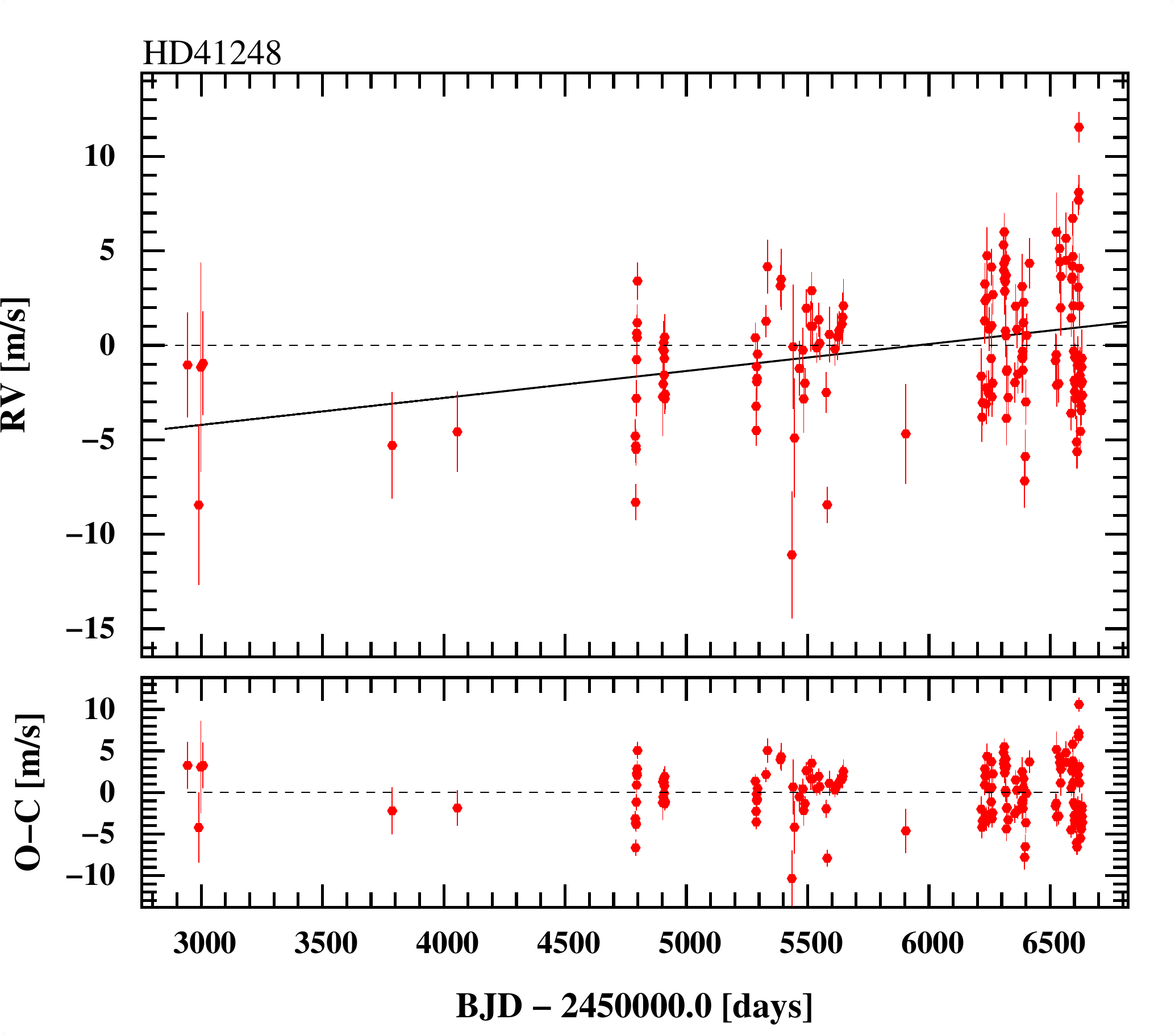}}
\caption{Time series of the radial velocity and a linear fit to the data. The residuals are shown in the lower panel.} 
\label{fig:rvtrend}
\end{figure}

In order to take the analysis of the data one step further, we defined three different sets: set\,\#1, which corresponds to the data that was used by Jenkins et al. (JD up to 55647), set\,\#2 with
JD between 55904.8 and 56414.5, and set\,\#3 with JD between 56521.9 and 56632.7. Sets\,\#2, and \#3 correspond to two different observing seasons, and are separated by a temporal gap (due to
the passage of the star ``close'' to the Sun). The number of points in each data set is 61, 50, and 45, respectively for set\,\#1, \#2, and \#3. In the following, and before dividing the data in the three different sets, we fitted and subtracted from the RVs a linear trend (Fig.\,\ref{fig:rvtrend}), 
with a slope of 0.52$\pm$0.14 m\,s$^{-1}$/yr. Since the signals we are exploring in this paper are all of relatively short period, this decision will not have an impact on the presented results.

\begin{figure}[th!]
\resizebox{\hsize}{!}{\includegraphics[]{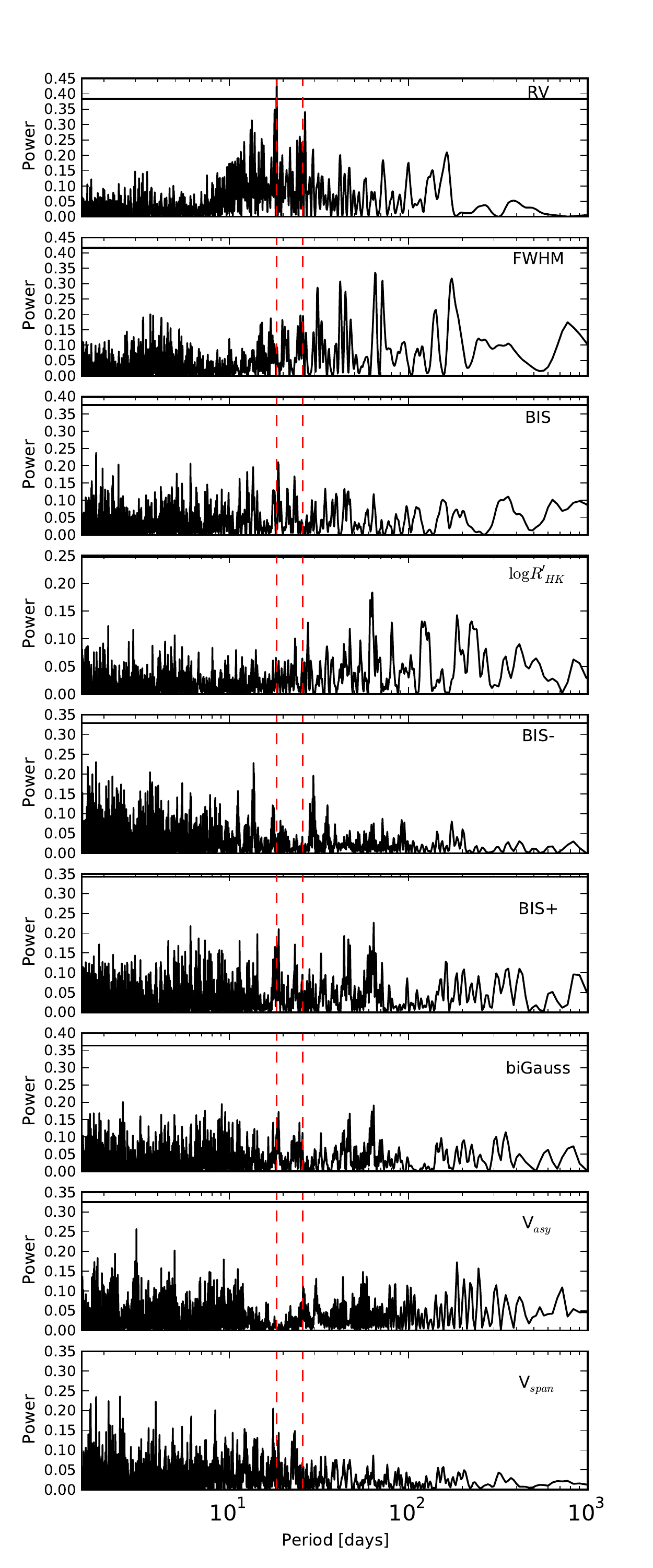}}
\caption{Periodograms of (from top to bottom) the RV, FWHM, BIS, $\log{R'_{HK}}$, BIS$-$, BIS$+$, biGauss, V$_{asy}$, and V$_{span}$ for data set\,\#1. The horizontal line denotes
the 1\% false alarm probability level. Vertical dashed lines denote the position of the 18.36 and 25.7\,day signals as found by \citet[][]{Jenkins-2013}.}
\label{fig:perioset1}
\end{figure}

\begin{figure}[th!]
\resizebox{\hsize}{!}{\includegraphics[]{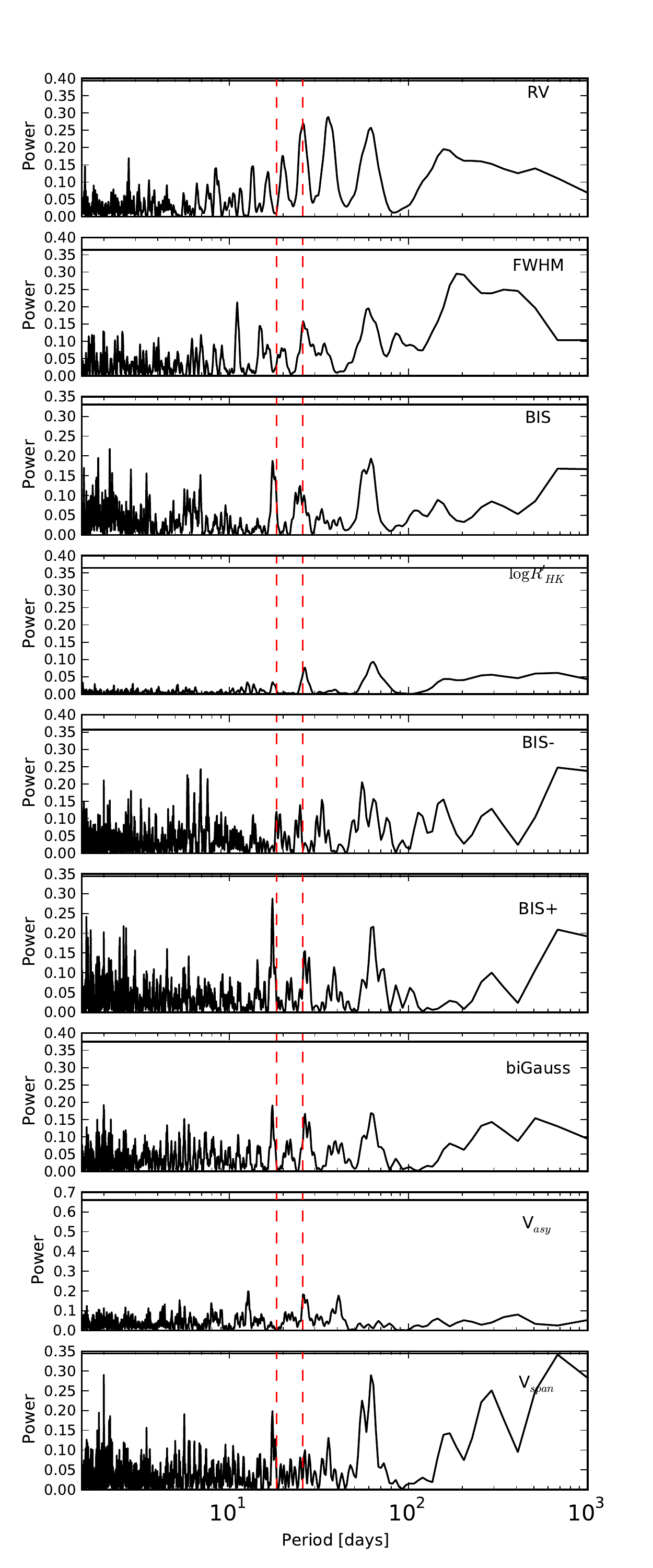}}
\caption{Same as Fig\,\ref{fig:perioset1} for data set\,\#2.}
\label{fig:perioset2}
\end{figure}

\begin{figure}[th!]
\resizebox{\hsize}{!}{\includegraphics[]{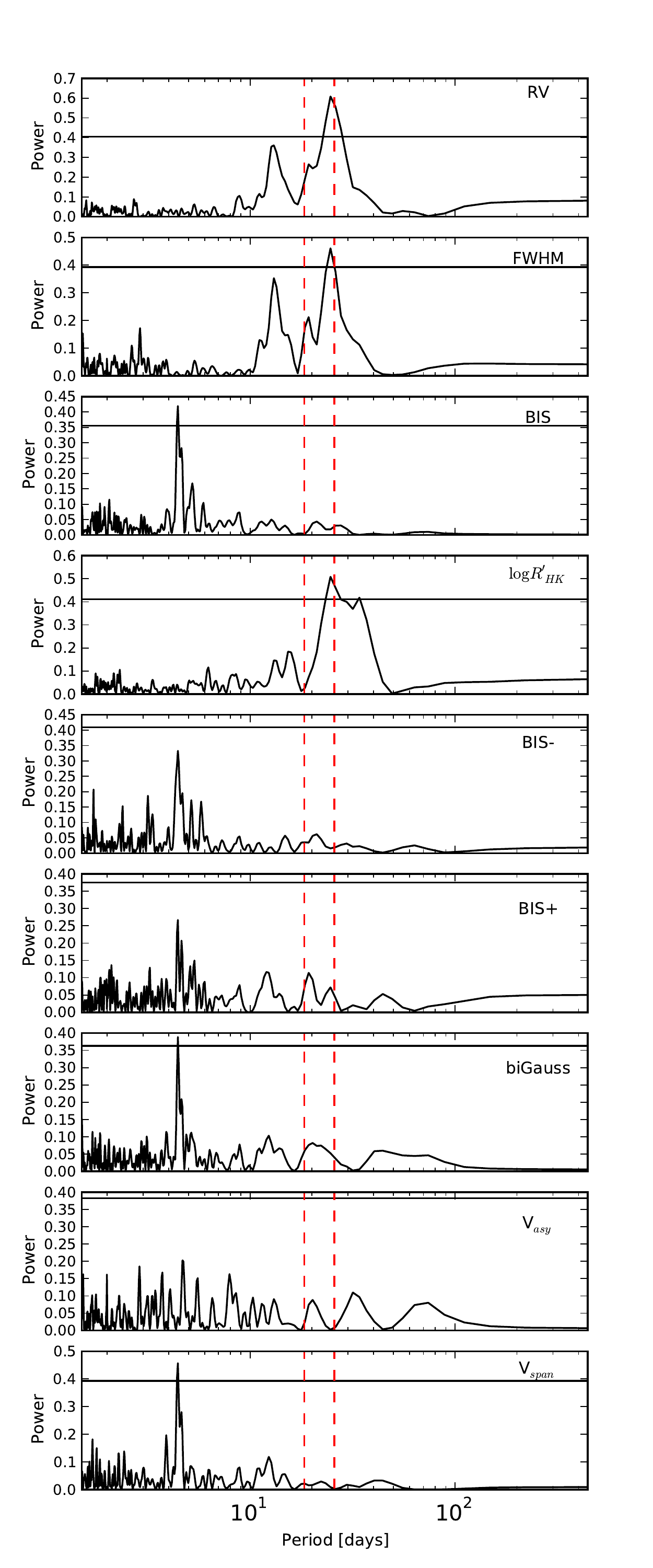}}
\caption{Same as Fig\,\ref{fig:perioset1} for data set\,\#3.}
\label{fig:perioset3}
\end{figure}

\begin{figure}[th!]
\resizebox{\hsize}{!}{\includegraphics[]{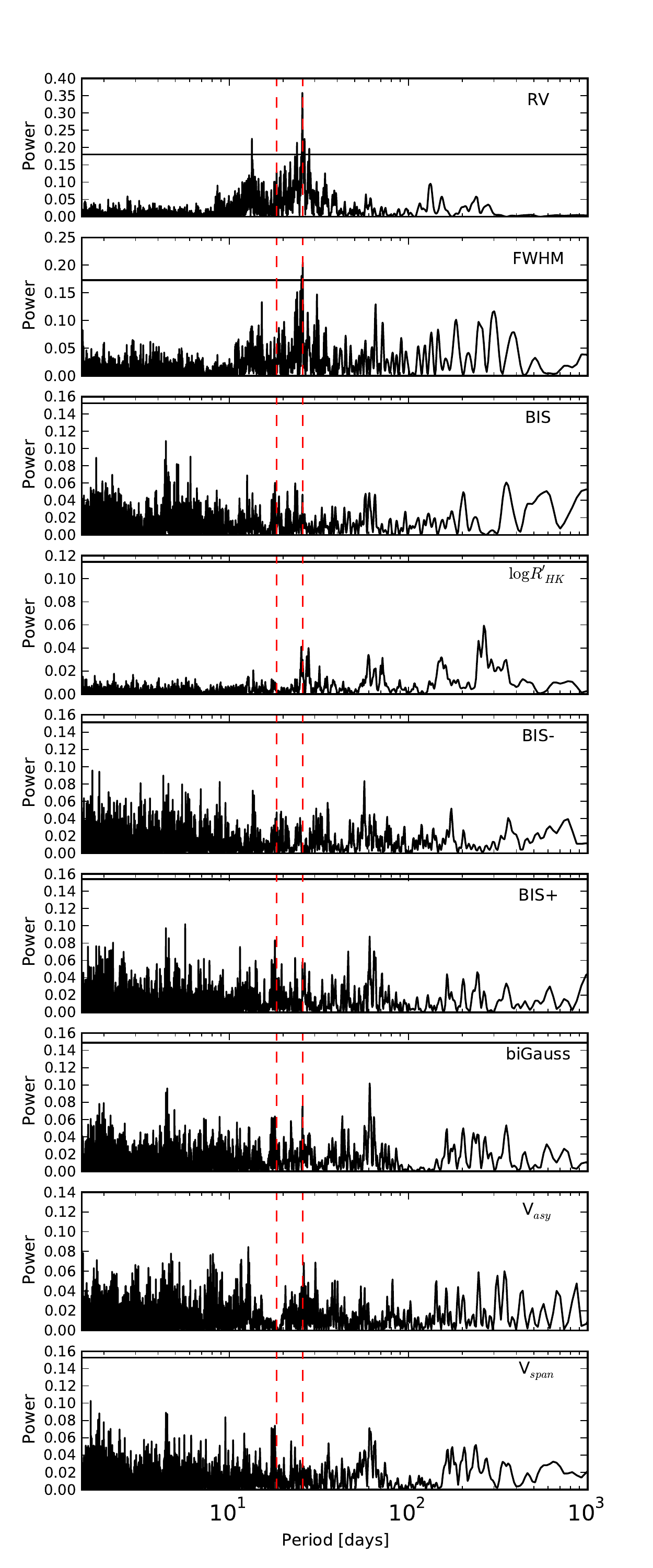}}
\caption{Same as Fig\,\ref{fig:perioset1} for all the data.}
\label{fig:perioset123}
\end{figure}

In Figs.\,\ref{fig:perioset1}, \ref{fig:perioset2}, \ref{fig:perioset3}, \ref{fig:perioset123} we present, from top to bottom, the GLS of the RV, FWHM, BIS, $\log{R'_{HK}}$, BIS$-$, BIS$+$, biGauss, V$_{asy}$, and V$_{span}$ for the data in sets \#1, \#2, \#3, as well as for all our data together, respectively. In all the plots, the horizontal line denotes the 1\% False Alarm Probability (that we will consider as the significance limit). This value was computed using a permutation test; more details can be found in \citet[][]{Mortier-2012}. The vertical dashed lines
denote the locus of the 18- and 25-day periods. 

In set\,\#1 (Fig.\,\ref{fig:perioset1}), the highest and only significant peak in RV is at $\sim$18\,days, as already pointed out in \citet[][]{Jenkins-2013}. 
This peak corresponds to the signal that these authors attributed to the presence of one of the planets.
No similarly significant peak is seen in any of the other indices, though
a clear peak near 18\,days is also observed at least in the BIS and BIS$+$ line profile indicators. The peak at $\sim$25\,days, which
corresponds to the second candidate planet announced by Jenkins et al., also has some power in
the FWHM and V$_{asy}$, though never at a significant level. A peak close to 25\,days is also present
in $\log{R'_{HK}}$. Finally, a peak close to 30\,days (one of the periods mentioned in Sect.\,\ref{sec:bayes}) is also seen in FWHM, BIS$-$, and V$_{asy}$.

For data set\,\#2 alone, the periodograms, presented in Fig.\,\ref{fig:perioset2}, show that no significant peaks
are detected in any of the variables. In RV, a forest of peaks is present, the most conspicuous a peak at $\sim$35\,days (with some power also
at similar value in FWHM, BIS$+$, and V$_{asy}$), followed by
the one at $\sim$25\,days. This signal, at or close to 25\,days, is also seen in all the remaining variables, with exception of the V$_{span}$. 
For RV, no peak is present at 18\,days, though clear peaks close to that period are observed in BIS, BIS$+$, biGauss, and V$_{span}$. 
Finally, a clear peak at $\sim$60\,days is seen in all the variables analyzed, with exception of V$_{asy}$.

For data set\,\#3, Fig.\,\ref{fig:perioset3} shows that for RV, FWHM, and $\log{R'_{HK}}$ there is a clear signal at
25\,days, as well as at its first harmonic (P/2$\sim$12.5). No clear signal is observed at 18.36\,days, though a non-significant bump
in the peridogram exists at $\sim$19\,days. A peak at $\sim$19\,days is also seen in the periodogram of the FWHM.
Interestingly, the periodogram of BIS, BIS$-$, BIS$+$, biGauss, and V$_{span}$ shows the presence of a significant peak at $\sim$4.5 days, 
a value that is 1/4th of 18\,days. The cause for this peak will not be discussed further as we don't have any clear 
explanation for its existence. 

Finally, the periodograms of the whole data set (Fig.\,\ref{fig:perioset123}) show, as already mentioned above, that the pattern
observed in RV is also well reproduced in the FWHM, with the clear and significant 25\,day period signal present
in both variables (the first and second harmonics, P/2 and P/3, are also visible
at least in the RV). No peak at 18\,days is seen in the periodograms, though a hint
of power at $\sim$19.5\,days is seen in the FWHM. The GLS of BIS, BIS$+$, biGauss, and V$_{span}$ also shows some power close to 18\,days, but 
no significant peak is seen. A peak around 31\,days is also observed in the FWHM and in V$_{asy}$.

Noteworthy also is the fact that the amplitude of the $\sim$25-day period signal seems to increase
as we move from set\,\#1 to set\,\#3. The analysis of the RV and $\log{R'_{HK}}$ periodograms further shows that its phase did not significantly vary over time. 
In complement with the analysis presented below, this suggests that we may be in the presence of
a signal produced by an evolving (growing) active longitude \citep[][]{Berdyugina-2003,Ivanov-2007}, that kept its position in the stellar disk 
approximately constant over the last years.

In brief, the periodogram of the RV shows a complex pattern, that clearly evolves as a function of time,
rendering our analysis of the data complex and difficult. The same is true for the activity and line-profile indicators.
The periods found by the Bayesian fitting procedure mentioned above, for example, are all correspondent with 
peaks in stellar activity or line profile indicators, varying over time. 

We should note that in a case where multiple signals (e.g. red noise or other Keplerians) with high enough, significant amplitudes are present in the data, one coherent
signal may in principle remain undetected by a periodogram analysis, even if it is still present in the data (since it will may be diluted by the remaining signals).
This problem is actually present in any analysis of data in the absence of a full model.
Several of the tests presented here and in the following sections are thus valid under the assumption that no additional,
sufficiently stronger signals exist that can (at least completely) hide the periodicities we are testing. Note however that all our analysis
is based on several diagnostics, rendering it more solid.

\subsection{The 25-day period}

As discussed in Sect.\,\ref{sec:periodograms}, from the analysis of the whole data set there is no sign of the 18-day period signal clearly observed in \citet[][]{Jenkins-2013}. However, a distinctive peak at $\sim$25 days dominates the GLS. A second and third peak, at about $\sim$13 and $\sim$8.5-days are also observed. These two peaks are at the approximate position of the first and second harmonics of the $\sim$25-day period.

The periodogram can be interpreted in the light of at least three distinct scenarios: i) the observed signal is induced by the presence of one eccentric 
planet (as fitted in Sect.\,\ref{sec:keplerians}) with a period of 25-days (whose Keplerian signal produced a periodogram showing the periods and its harmonics), ii)
we are in the presence of a system of several planets with periods that are in resonance with the 25-day period signal (see Sect.\,\ref{sec:bayes}), or iii) we are in the
presence of a signal induced by stellar spots or other activity related phenomena. The pattern observed
is indeed very similar to the expected RV signal caused by a spotted star as presented and discussed in \citet[][]{Boisse-2011}.

\begin{figure}[t!]
\resizebox{\hsize}{!}{\includegraphics[viewport=10 150 580 720]{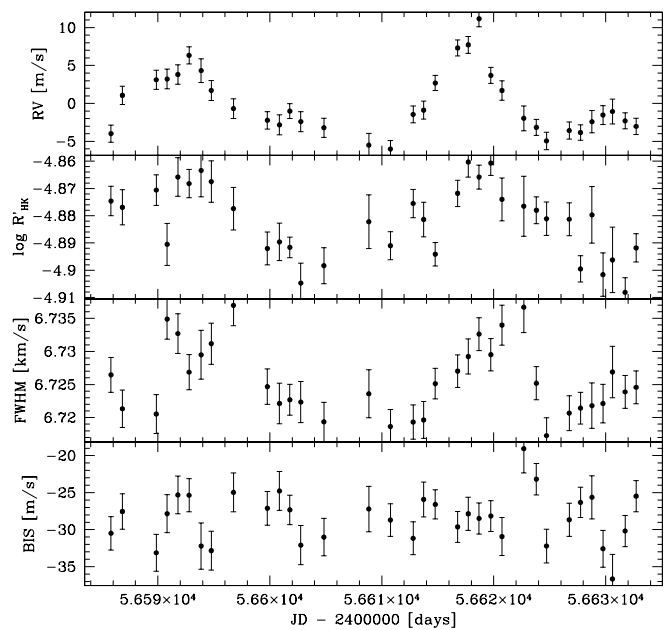}}
\caption{Time series of the radial velocity, stellar activity, FWHM, and BIS for the period between JD= 2456580 and 2456640. }
\label{fig:lastdata}
\end{figure}

In Fig.\,\ref{fig:lastdata} we present the time series of the RV, $\log{R'_{HK}}$, FWHM, and BIS for the last series of RV data, obtained in the end of 2013 (corresponding to the last measurements of set\,\#3). As we can see from the plot, there is a clear correlation between the RV and both the FWHM and the stellar activity index $\log{R'_{HK}}$. With this information we conclude that the 25-day period signal observed in RV (and its harmonics) most likely corresponds to the rotational period of the star, and that the RV signal observed is caused by the rotational modulation of activity features on the stellar photosphere. The 25-day signal announced by \citet[][]{Jenkins-2013} is thus most likely better explained by stellar activity and not by the existence of a planet orbiting HD\,41248.

We should note that the analysis of the GLS periodograms shows that the phase of the RV signal plotted in Fig.\,\ref{fig:lastdata} (and in particular the peak in its value) is about 
$\sim$35 and $\sim$15 degrees behind the one observed in FWHM and $\log{R'_{HK}}$, respectively. 
This lag is expected if the RV signal is induced by stellar spots \citep[see e.g.][]{Forveille-2009}. Indeed,
when the active regions are appearing and occupy the blueshifted side of the star, the RV will show an increasing value with time.
Simulations with the SOAP code \citep[][]{Boisse-2011}\footnote{http://www.astro.up.pt/soap} show that the maximum of this RV will occur when the spots 
are $\sim$45 degrees from ``meridian'' (or close to disk center). Given the simple physics\footnote{The present SOAP version does not 
include, e.g., the modeling of convective blueshifts, which are different in active regions.} used in the model, we consider that 
this number is compatible with the observed value. The value will then decrease to zero when the spot is at ``meridian''. 
This instant sets the maximum activity level as the active region shows its maximum projected area\footnote{This would also correspond 
to the maximum photometric variability.}.

{
As a complementary test, we computed the Pearson's correlation coefficient ($\rho$) between the RV and the different line profile indicators for data set\,\#3. 
For the correlation with FWHM, a value of $\rho$=0.52 was obtained. A Monte Carlo simulation was then done to calculate the probability of reaching this value due to a change alignment of the data points. 
This test was done by performing a Fisher-Yates shuffling of the values of RV and FWHM 100\,000 times, computing the correlation coefficient for each simulated dataset, and deriving the distribution of the resulting $\rho$ values. For details about the method and its background we point to \citet[][]{Figueira-2013}. The test showed that the observed $\rho$ is at 4.5 sigma from an uncorrelated (shuffled) distribution, meaning that it is very unlikely that it is caused by a chance event. Note that despite the significant correlation found, the value of $\rho$ is not particularly high. This is due to the fact that the RV and the different line profile indicators are usually not correlated with a 1:1 relation, among other reasons \citep[see e.g.][]{Figueira-2013}. This point is also illustrated by the phase shift observed between the FWHM and RV as discussed above.
}

\subsection{The 18-day period}
\label{sec:18period}

The result presented above does not {\it per se} discard the presence of planets orbiting HD\,41248 at other periods. In particular,
they do not allow us to discard the existence of an 18-day period signal as present in the first batch of data 
and interpreted by \citet[][]{Jenkins-2013} as caused by the presence of a super-Earth mass planet.

As discussed above, however, if we divide the whole data set in three different groups, the GLS analysis suggests that 18-day period signal is 
only observed in the first dataset (set\,\#1), which corresponds to the data used by Jenkins et al. No signature of the 18-day period
is visible in the remaining data, even if the number of points in sets\,\#2 and \#3 are similar to the ones in set\,\#1. 
One can then ask if the 18\,day period signal is still present in the data, i.e., if it is constant
over time, or alternatively if it was only present in the first data set. To address this issue we decided to make a series of tests, as follows.

\subsubsection{Subtracting the 25-day period signal}
\label{sec:harmonic}

Set\,\#3 has, by far, the best time coverage of the data. That makes it particularly suitable
to analyze the existing signals. 
We thus decided to analyze this set in detail to test if the 18-day period signal can be retrieved
after removing the signal at 25-days.

To do this we used the approach of \citet[][]{Boisse-2011} to fit the rotational period
and its harmonics, as successfully done by \citet[][]{Dumusque-2012} for the case of $\alpha$\,Cen\,B. Heretofore we applied two methods. 
In the first one, we fitted a P$\sim$25\,day sinusoidal, together with the first harmonic (at P/2) to
the RV time series. In the second case we decided to use the FWHM variation as a proxy for activity-induced RV variations,
and fix P using the analysis of the FWHM signal. This procedure
allows us to guarantee that in the fitting process we are not absorbing signals
that are e.g. present in the RV but are not induced by activity related phenomena (e.g. real planetary signals). The residuals
of the fit using both methods were analyzed.

\begin{figure}[t!]
\resizebox{\hsize}{!}{\includegraphics[]{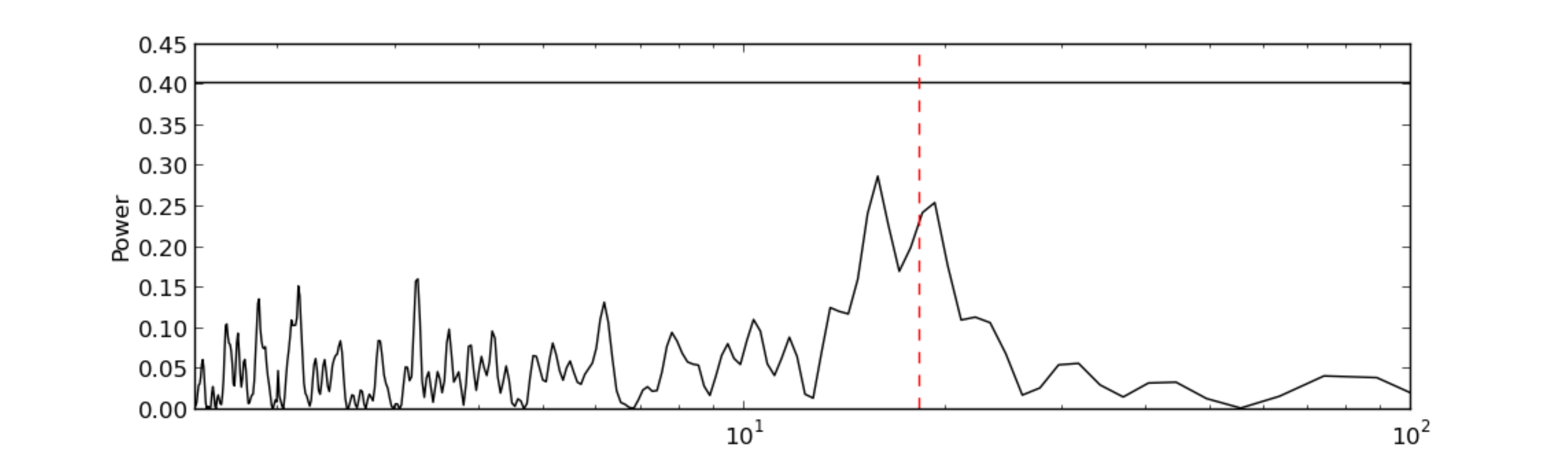}}
\resizebox{\hsize}{!}{\includegraphics[]{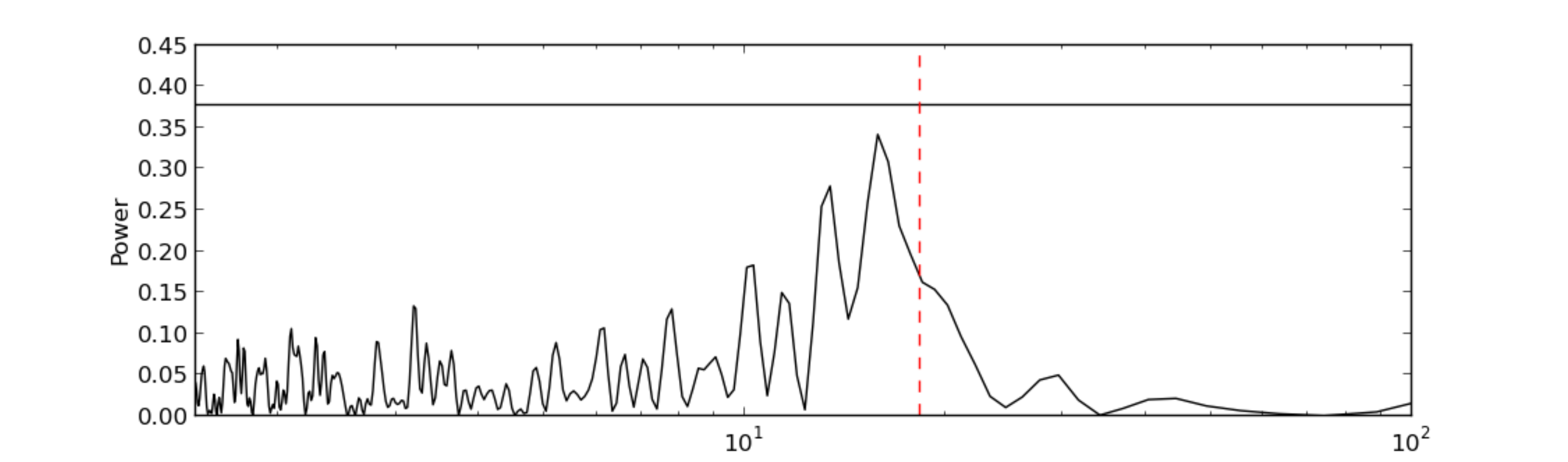}}
\caption{Periodograms of the residuals of the harmonics fit described in Sect.\,\ref{sec:harmonic}, making use of the RV
(top) and FWHM (bottom) to fix the period used to subtract the RV signal and its harmonics. The dashed line indicates
the position of the 18.36\,day period. The horizontal line represents the 1\% FAP.}
\label{fig:harmonic}
\end{figure}

The results of both tests can be seen in Fig.\,\ref{fig:harmonic}, where we present the GLS of the residuals
to both fits. In both panels, the dashed line indicates the 18.36\,day period, while the horizontal line represents the 1\% FAP.
As can be seen in the plots, no significant peak exist at $\sim$18\,days. The highest peak in each plot is
at 15.77 and 15.94\,days, respectively. The second highest signal in the top panel is at a period of 19.02.

Although not conclusive, this test does not lend support to the planetary explanation for the 18.36-day signal presented 
in \citet[][]{Jenkins-2013}. In fact, as we will see in the next section, if that signal was present in data 
set\,\#3 it should have been easily spotted. We assume here that no significantly stronger signal at other period 
was present that could ``mask'' it. The absence of any strong peak in the GLS of Fig.\,\ref{fig:harmonic} at any
other period gives support to this assumption.

\subsubsection{Simulating the data}

As a second test, for each of the 3 time series of data mentioned above, we generated a set of full synthetic radial velocities.
The data was generated considering the real observed dates to mimic the real time sampling. The error bars for each RV point were also kept as in the original data. 
On each data set, we first added a Keplerian signal of $\sim$18.36-day period as observed by \citet[][]{Jenkins-2013}. White noise 
was then added to each point in agreement with the error bars to simulate the different average measurement errors in each data set.

\begin{figure}[t!]
\resizebox{\hsize}{!}{\includegraphics[bb = 10 20 450 550]{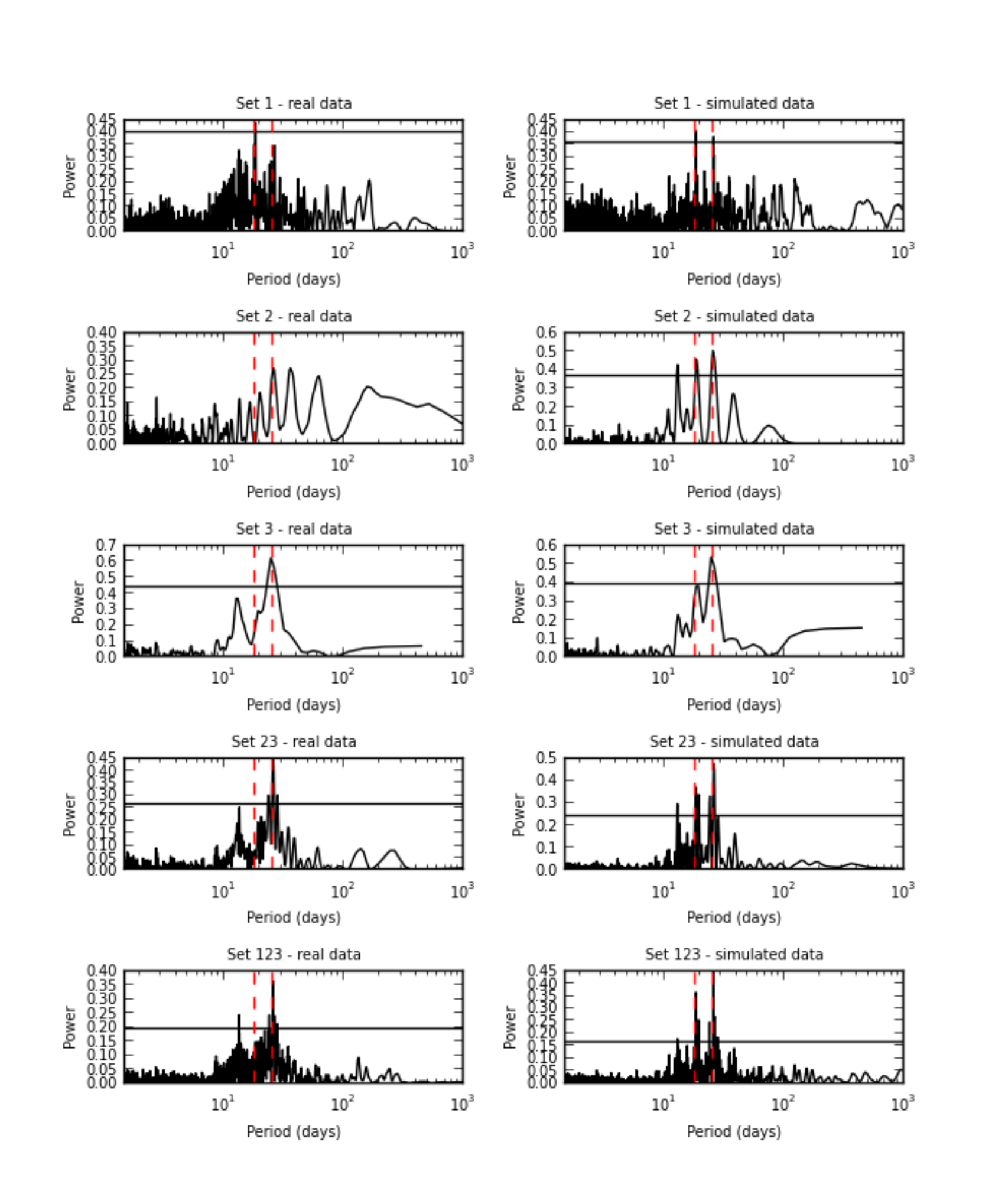}}
\caption{Periodograms of the different data sets, both of the real data (left) and simulated data (right). Vertical dashed lines represent the position
of the 18- and 25-day signals presented in Jenkins et al.}
\label{fig:peridograms}
\end{figure}

On top of this we injected a signal of 25-days into the data, as fit to the last data set where the time coverage of the
data clearly allows us to model the 25-day signal. Again, the fit was done using the approach in \citet[][]{Boisse-2011}, i.e.,
fitting the rotational period and its first harmonic (P/2)\footnote{The third harmonic, or P/3, did not present a significant power.}.
On each of the 3 datasets (set\,\#1, \#2, and \#3), however, the injected 25-day period signal was varied in amplitude until the rms of the synthetic data
is the same as measured in the real data. 

Note again that we find evidence that the observed activity signature has been growing over time. Not only do
the peridograms show that the 25-day peak increases its significance from set\,\#1 to set\,\#3, but
the rms of the data also increased from 2.6\,m\,s$^{-1}$ in set\,\#1 to 3.0\,m\,s$^{-1}$ in set\,\#2, and finally to 3.2\,m\,s$^{-1}$ in set\,\#3.

In Fig.\,\ref{fig:peridograms} we present both the observed and simulated peridograms of the data
for the whole data set (set\,\#123), as well as for set\,\#1, set\,\#2, set\,\#3, and for sets\,\#2 and \#3 together (set\,\#23). As we can see from the plots, while
in the simulated data the 18-day period signal was always clear (even if often with an amplitude lower
than the one seen for the 25-day period signal), the situation in the real data is different: except for the first set of data (set\,\#1),
the 18-day period signal is not observed in any other data set. In other words, the simulations presented
here suggest that the signal at 18\,days should in principle have been clearly detected in
sets\,\#2 and \#3 if it had the same amplitude and phase as found in set\,\#1. Once again this result does not support the 
scenario of the existence of an 18-day period signal as reported in \citet[][]{Jenkins-2013}.

\subsubsection{Bayesian analysis including activity}

To further test if the 18-d signal is supported by the new data, we performed a new Bayesian analysis following the same procedure as in Sect.\,\ref{sec:bayes} but using data set\,\#3 alone. 
This time, however, we modeled the 25-day period activity signal as in Sect.\,\ref{sec:harmonic}, using two sines at P$_{rot}$ and P$_{rot}$ /2 \citep[][]{Boisse-2011}.

We then computed the Bayes factor between the following two models: an activity signal at $\sim$25d with a Keplerian at $\sim$18\,days and a 
$\sim$25d activity signal alone. The results indicate that, statistically, we cannot distinguish between the two models.
This therefore strongly suggests that the 18-d planet, as found by \citet[][]{Jenkins-2013}, is not confirmed (though also not rejected) by the new
observations of dataset \#3.


\section{Analyzing the residuals: planet detection limits}

Assuming that the 18- and 25-day signals detected in set\,\#1 are purely of stellar origin (induced
by activity), we can test if any other signal exists in the data that can be attributed to
a planet. 

As a first note of caution, its important to note that at present we do not have the
necessary tools to model the whole dataset in a correct, physical way. This is because the activity
pattern in HD\,41248 has been shown to be complex, inducing clear but variable signals in
amplitude and (likely) in period as a function of time. No strictly periodic signal is thus valid
when modeling the whole data, independently of the methodology used for the fit (e.g. frequentist analysis vs. bayesian fittings).
This implies that we cannot simply model the whole data set with e.g. a series of Keplerian functions.

To test the existence of further signals, we then first removed the two signals present in the first data
set by fitting a 2-Keplerian function. The best fit found is similar to the one derived by
\citet[][]{Jenkins-2013}, though in our case we found an eccentricity of 0.38 for
the 18-day period Keplerian fit. Note that the 25\,day period signal is not
statistically significant in data set\,\#1. However, since it has been shown to be coherent and have the same
origin as the clear signal found in set\,\#3, we decided to remove it.

For set\,\#2, since no significant peaks appear in the RV periodogram, we have not removed any signal. 
For set\,\#3 we again removed the 25-day signal and its first harmonic as discussed in Sect.\,\ref{sec:harmonic},
while fixing P using the RV dataset itself. 

\begin{figure}[t!]
\resizebox{\hsize}{!}{\includegraphics[bb = 100 100 650 500]{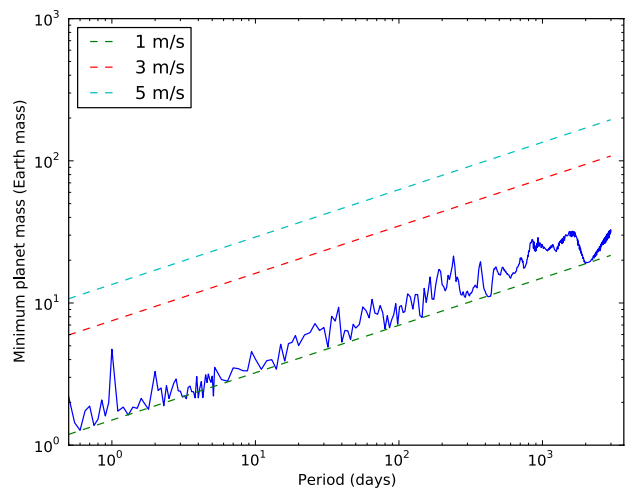}}
\caption{Minimum planetary mass against period. The solid line
represents the detection limits. The dashed lines indicates
a circular planetary signal with RV semi-amplitude of 1, 3 and 5 m\,s$^{-1}$.}
\label{fig:limits}
\end{figure}

After removing the different signals in the 3 different sets we analyzed the joint data using the GLS. The results show that the highest peak appears at 193\,days, but a
permutation test shows that it is not significant (it has a False Alarm Probability of 30\%).

With this set of residuals we could also derive the detection limits of potential planets
present in the data. For this we used the same approach as in \citet[][]{Mortier-2012}.
The results of this analysis are presented in Fig.\,\ref{fig:limits}, and suggest that
we can reasonably exclude any planets with mass above 10\,M$_\oplus$ in
the period range up to $\sim$100\,days. This value decreases to $\sim$4M$_\oplus$
if we restrict the period range to below 10\,days.

Note that the results do not significantly vary when you use the FWHM to fix P when removing the
signal in set\,\#3. Also, no significant differences are observed if we only subtract the 
18-day period signal in set\,\#1.


\section{Discussion and conclusions}
\label{sec:conclusions}

In a recent paper, \citet[][]{Jenkins-2013} reported the existence of a system of two low-mass
planets orbiting HD\,41248 in almost circular orbits of periods $\sim$18 and 25\,days.
In this paper we analyzed this system after adding almost 160 new radial velocity points obtained
with the HARPS spectrograph.

The results of this analysis do not allow us to confirm the planetary origin of 
the signals observed in the RV data of HD\,41248 as previously suggested by \citet[][]{Jenkins-2013}. The observed
25-day period signal is almost exactly reproduced in the stellar activity
index $\log{R'_{HK}}$ as well as in the FWHM of the HARPS CCF. This signal has a complex structure and varying 
amplitude with time, making it difficult to model with present day tools. This fact renders
the analysis of the putative 18-day periodicity difficult. However, although we cannot fully discard the existence of a stable, 
periodic signal at 18\,days as expected from the presence of a planet, the different tests that we conducted show that the current data (both the RV and activity/line profile indicators) 
does not support its existence. In brief, the 25-day period signal detected by \citet[][]{Jenkins-2013} is best explained as induced by stellar activity phenomena.
Our analysis also suggests that the 18-day signal may have a similar origin.

We assume here that at a period of 25-days, a Neptune like planet will not be able to induce strong tidal or magnetic
interactions with the star, which could result in an activity signature with a period similar to the orbital period
of the planet \citep[][]{Saar-2001,Shkolnik-2003}\footnote{Or possibly half the orbital period in case of tidal interaction.}.
We note that cases have been found where the orbital period seems to coincide, within the uncertainties, with the rotational period
of the host star \citep[][]{Santos-2003b}. If this is the case for
HD\,41248, the low amplitude of the signals and the complexity of the data will make it very
difficult to confirm.

The complexity of the signals observed and the estimate for the rotational period of the star ($\sim$20\,days -- Table\,\ref{tab:parameters})
leads us to propose that the observed 18\,day and 25\,day signals may be caused by at least two different
active regions/longitudes in a star presenting a strong differential rotation pattern.
In this scenario, the 18 and 25\,day period signals would imply a differential rotation with an amplitude of about 25\%. 
The Sun itself rotates, at the equator, with a rotational period of 26\,days, while at the poles the
value increases to $\sim$35\,days. Higher levels of differential rotation have been found in earlier
type stars \citep[][]{Barnes-2005,Reiners-2006,Ammler-2012,Reinhold-2013}\footnote{\citet[][]{Gastine-2014} suggest that
the cooler stars may even present antisolar differential rotation, where the poles rotate faster than the equator.}.
A difference in rotational period of 25\% in the surface of HD\,41248 seems thus perfectly plausible.
This scenario would explain the existence of a growing 25-day period signal, caused by
a growing active region that kept its phase all over the period of our measurements, as well as the disappearance of the 18-day period signal, if caused
by an active region that disappeared (or became much weaker) and was positioned at a lower stellar latitude. It would also provide 
a simple explanation for the forest of peaks observed in data set\,\#2, if we assume that
other active regions may have appeared and disappeared at other latitudes.

One alternative scenario to explain the observed complex pattern is related with the
fact that the data presented above present a very complex structure. It is clear from the plots that
the activity patterns we are observing in this star present signatures of having evolved
over the timespan (more than 10 years) of our measurements. An interesting hint may come, however, from the study of \citet[][]{Lanza-2003} where the authors 
analyzed the rotational period of the sun using the Total Solar Irradiance (TSI) observed during the maximum of the eleven-year cycle.
In the Sun, large spot groups have typical lifetimes of 10-15 days, while the rotational period is close to 25\,days. The fact
that the timescales for spot evolution are shorter than the rotational period, together with the appearance
and disappearance of new spot groups in different rotational phases, renders the derivation of rotational
periods (from the data) a complex issue. As a result, \citet[][]{Lanza-2003} have found that, during the 1999-2001 period
when the Sun was close to solar maximum, it was impossible to properly retrieve the rotational period of the Sun
using the TSI data, as the analysis yielded values from 24 up to 31\,days. Given the complex pattern of
data presented in the present paper for HD\,41248, together with the uneven sampling, the
presence of signals at 18 and 25 days may simply reflect a difficulty in fitting the data properly (at least using ``simple'' Keplerian functions). 

The present paper presents a good example of how difficult the analysis of radial velocity data can be when
searching for very low-mass planets that induce low-amplitude signals, close to the measurement precision.
The results also point very clearly the importance of following a star for a sufficiently long period of time 
until one can confidently secure the characterization of the whole system, including the effects
of stellar activity. In this particular case, a proper sampling of the data (as in set\,\#3) was fundamental to disentangle the
sources of the radial velocity signals. 

This study also shows that Bayesian analysis are not immune from false-positive detections, especially in the presence of stellar activity which might not be approximated by a 
series of Keplerian functions. The present case also demonstrates how important it is to make use of methodologies and tools to model and understand the signals produced by 
stellar activity. A complete characterization of the data may imply the development of more detailed physical models of stellar activity and
its impact on radial velocity measurements \citep[e.g.][]{Boisse-2012}, as well as of more sensitive diagnostic methods \citep[e.g.][]{Figueira-2013}. 
Without that it will be very difficult to fully analyze these systems with any statistical/fitting procedure. The amplitudes of the RV signals 
imposed by stellar activity are, even in the case of a relatively inactive
star such as HD\,41248, often of the same order of magnitude as the expected signals due to a low mass planet. Alternatively,
complementary spectroscopic measurements using other wavelengths (e.g. near-IR) may be useful to disentangle real planets
from activity induced signals \citep[e.g.][]{Huelamo-2008,Prato-2008,Figueira-2010b}. A new generation of near-IR spectrographs is presently being 
developed \citep[e.g. CARMENES and Spirou --][]{Quirrenbach-2014,Delfosse-2013}, opening great perspectives in this domain.

\begin{acknowledgements}
We would like to thank N. Lanza for the fruitful discussions. We acknowledge support from Funda\c{c}\~ao para a Ci\^encia e a Tecnologia (FCT, Portugal) through FEDER funds 
in program COMPETE,  as well as through national funds, in the form of grants reference PTDC/CTE-AST/120251/2010 (COMPETE reference FCOMP-01-0124-FEDER-019884), RECI/FIS-AST/0176/2012 (FCOMP-01-0124-FEDER-027493), and RECI/FIS-AST/0163/2012 (FCOMP-01-0124-FEDER-027492). 
This work was supported by the European Research Council/European Community under the FP7 through Starting Grant agreement number 239953.
NCS and PF were supported by FCT through the Investigador FCT contract references IF/00169/2012 and IF/01037/2013 and POPH/FSE (EC) by FEDER funding through the program "Programa Operacional de Factores de Competitividade - COMPETE. X. Dumusque was supported by the 
Swiss National Science Foundation. This work made use of the SIMBAD database.
\end{acknowledgements}


\appendix
\section{Line Profile Analysis Suite}
\label{appendix}

In a recent paper, \citet[][]{Figueira-2013} analyzed a series of line profile indicators and discussed how these can help us pinpoint a RV signal 
created by stellar phenomena. These indicators and the associated statistical tests are now wrapped up in a simple code made available in the ExoEarths software 
webpage\footnote{\url{http://www.astro.up.pt/exoearths/tools.html}}. A living version of it can be accessed through a bitbucket repository\footnote{\url{https://bitbucket.org/pedrofigueira/line-profile-indicators}}.


The program is run simply by calling it using python (i.e. {\it python LineProf.py}), with all the information being provided by an ASCII configuration file. No programming experience is thus required; we note however that the program was written in a modular way, so that it can be used as a building block for complex data analysis software. 

The program reads automatically a list of FITS files (e.g. HARPS-N or HARPS-S), or ASCII data with the CCFs to analyze. It applies the indicators presented in \citet[][]{Figueira-2013}, and evaluates the correlation between such indicators and RVs. Then 100 000 non-correlated data sets are obtained by doing a Fisher-Yates shuffle of the data pairs, and the correlation of the original set compared with the correlation of the shuffled set. The z-value is provided, along with the (Gaussian) probability that the correlation is drawn from an uncorrelated data set. All these results are stored in ASCII files, and paper-quality plots for all the indicators selected are generated. The program can digest several dozens of files and do the complete analysis in a couple of minutes on a normal desktop/laptop.

\end{document}